\documentclass[aps,prc,showpacs,twocolumn,superscriptaddress,amssymb]{revtex4}
\usepackage{graphicx}
\def\la{\langle}
\def\ra{\rangle}
\def\beq{\begin{equation}}
\def\eeq{\end{equation}}
\def\be{\begin{eqnarray}}
\def\ee{\end{eqnarray}}

\def\k2av{\la k_\perp^2\ra}

\newcommand{\dd}{ {\textrm d}}

\begin{document}
\title{Intrinsic parton transverse momentum in next-to-leading-order 
pion production}
\author{G\'abor Papp}
\email{pg@ludens.elte.hu}
\affiliation{Department for Theoretical Physics,
E{\"o}tv{\"o}s University, \\ P{\'a}zm{\'a}ny P. 1/A, Budapest 1117, Hungary}
\author{Gergely G. Barnaf\"oldi}
\email{bgergely@rmki.kfki.hu}
\affiliation{RMKI KFKI, P.O. Box 49, Budapest 1595, Hungary}
\author{P\'eter L\'evai}
\email{plevai@rmki.kfki.hu}
\affiliation{RMKI KFKI, P.O. Box 49, Budapest 1595, Hungary}
\author{George Fai}
\email{fai@cnr4.physics.kent.edu}
\affiliation{CNR, Kent State University, Kent OH 44242, USA}

\date{\today}

\begin{abstract}
We study pion production in proton-proton collisions within a
pQCD-improved parton model in next-to-leading order augmented by
intrinsic transverse momentum ($k_\perp$) of the partons. 
We find the introduction of intrinsic transverse momentum necessary
to reproduce the experimental data in the CERN SPS to RHIC energy
range, and we study its influence on the so-called $K$ factor,
the ratio of the NLO cross section to the Born term.
A strong $p_T$ dependence is seen, especially in the 
$3-6$ GeV transverse momentum region of the outgoing pion, where nuclear
effects (e.g. the Cronin effect) play an important role.
\end{abstract}


\maketitle

\section{Introduction}
Today's collider facilities raise the interest in testing perturbative
QCD (pQCD) at work, and in searching for phenonema beyond its capability.
However -- at least for the experimentally available 
transverse momentum region -- pQCD parton models underestimate the
production of mesons in proton-proton ($pp$) collisions~\cite{LOBp,xnwang},
even at next-to-leading order (NLO)~\cite{aur1}. In order to restore the
consistency with the data, two methods were proposed at leading order
(LO): inclusion of the intrinsic transverse momentum of 
partons~\cite{LOBp,xnwang},
and/or an effective correction factor ($K$ factor), 
accounting for  higher order contributions~\cite{Eskola,Ramona}.
The physical background is to account for both, the
missing higher order perturbative corrections and radiation
effects. The latter is not present in
electron-proton processes, however, it becomes important
in $pp$ collisions~\cite{lai}.

In this paper we present the first results on pion production
in $pp$ collisions applying a NLO pQCD parton model {\em with}
intrinsic transverse momentum, taken from a
Gaussian distribution with width $\k2av$. In Section II the
intrinsic transverse momentum is introduced into the formalism, and the
appropriate NLO expressions are presented.
The necessity of such an extension is demonstrated at $\sqrt{s}=27.4$ GeV
in Section III. 
Next, we display the best fit values
of the width of the intrinsic transverse momentum distribution at NLO level
for available $pp\to\pi+X$ experiments, in the energy range $20$
GeV $\lesssim\sqrt{s}\lesssim 200$ GeV. Similarly to Ref.~\cite{LOBp},
we study the pion production in the $2$ GeV $< p_T < 6$ GeV transverse
momentum region, where nuclear effects are considered to be important.

Finally, we extract the ratio of the NLO cross section to the Born
term ($K$ factor) at different energies and transverse momenta, and
study its dependence on the amount of the included intrinsic
transverse momentum. This way we provide a numerical foundation to the
correction factors used in LO calculation~\cite{Wong}. 
We demonstrate, that a leading order calculation with a
fitted, c.m. energy, transverse momentum and scale dependent
$K$ factor and additional intrinsic transverse momentum
reproduces well the full NLO results at higher energies and momenta,
and can be used as a fast method to get a reasonable estimate of a full 
NLO calculation.

\section{Model}
In order to extend the applicability of the original, infinite momentum 
frame parton model~\cite{field} to smaller transverse momenta, 
we introduce the intrinsic transverse momentum of the partons~\cite{owens}. 
We write the four-momenta of the interacting partons ($a$ and $b$) 
as~\cite{Wong}
\be
 p_a &=& (x_a\frac{\sqrt{s}}2+\frac{k_{\perp,a}^2}{2x_a\sqrt{s}},\ 
        \vec{k}_{\perp,a},\ x_a\frac{\sqrt{s}}2-
        \frac{k_{\perp,a}^2}{2x_a\sqrt{s}})  \,\, , \\
 p_b &=& (x_b\frac{\sqrt{s}}2+\frac{k_{\perp,b}^2}{2x_b\sqrt{s}},\ 
        \vec{k}_{\perp,b},\
        -x_b\frac{\sqrt{s}}2+\frac{k_{\perp,b}^2}{2x_b\sqrt{s}}) \,\,
        .
        \nonumber
\ee
In this notation $x$, the momentum fraction carried by the
parton, becomes a parameter. The apparent fraction is $x-k_{\perp}^2/(x
s)$, however, for practical applications at high energy 
($p_T\gtrsim 3$ GeV; $\sqrt{s}\gtrsim 40$ GeV;
$\k2av\lesssim 2$ GeV${}^2$),
the distinction has a negligible ($\lesssim 5\%$) effect.
Furthermore, we require that the longitudinal direction of the partons 
does not change sign due to the transverse momentum, i.e. 
$x>k_\perp/\sqrt{s}$.


The starting point of our calculation is factorization: the hadronic
cross sections up to a power correction may be written as a convolution over
hard partonic (pQCD) processes,
\be
\label{pqcd-def}
        \frac{\dd\sigma}{\dd y\dd^2p_T} 
        \hspace*{-15mm}&\hspace*{15mm}=\hspace*{-15mm}&\hspace*{15mm} 
        \sum_{abc}
        \int\!\dd x_a\dd x_b\dd^2k_{\perp,a}\dd^2k_{\perp,b}\ 
        \frac{\dd z_c}{\pi z_c^2} \\
        & &f_{a/p}(x_a,Q;k_{\perp,a}) f_{b/p}(x_b,Q;k_{\perp,b}) 
        \frac{\dd\sigma}{\dd\hat{t}} D_{\pi/c}(z_c,Q_f) \,\, ,
        \nonumber 
\ee
where $\dd\sigma/\dd\hat{t}$ is the partonic cross section of
the reaction $a+b\to c+d$ (LO with condition
$\delta(1+(\hat{t}+\hat{u})/\hat{s})$), or $a+b\to c+d+e$ (NLO
with fixed $z_c$) and (at fixed scales) is the function of the 
partonic Mandelstam variables only. In order to avoid 
singularities due to the intrinsic transverse momentum we use regularization
\be
  \hat{s}\to\hat{s}+M^2, \quad \hat{t}\to\hat{t}-M^2/2, \quad
        \hat{u}\to\hat{u}-M^2/2 \,\, ,
\ee
with $M=1.8$ GeV.
The factorization is done at the {\em factorization} scale $Q$, where 
the parton content of the initial proton is determined
by the parton distribution
function $f_{a/p}$ (PDF). For simplicity, we assume a Gaussian dependence of
the PDFs on the intrinsic transverse momentum, with a width $\k2av$,
\be
  f(x,Q,k^2_\perp) = f(x,Q) \frac1{\pi\k2av} e^{-k_\perp^2/\k2av}.
\ee
We note, that such a separation may be viewed as a first approximation
to the {\em unintegrated} PDF, where recent studies indeed found a
shape close to Gaussian~\cite{updf}.
Finally, the hadrons are created collinearly with the outgoing parton
$c$ with momentum fraction $z_c$ at {\em fragmentation} scale $Q_f$.
The partonic cross section explicitly depends on three scales,
$Q$, $Q_f$ and the {\it renormalization} scale $Q_r$. 

In principle, the
scales can be determined such that the final result has the
minimum sensitivity to them~\cite{aur1}, and usually are set to be
equal. However, in this paper we use the results of a previous 
study~\cite{BP2002} of $pp$ and $pA$ data, where the reproduction the
Cronin effect in $pA$ reactions~\cite{Cronin} imposed such scales,
that the corresponding $\k2av\approx 2$ GeV$^2$.

We note that there is some ambiguity in the choice of scales. In the 
literature typical scales are fixed to hadronic 
or partonic variables, $\kappa p_T$ or $\kappa p_T/z_c$, respectively, where
$\kappa$ is an ${\cal O}(1)$ number. 
Other choices are also possible, e.g. an invariant scale,
$Q^2=\kappa^2\,\hat{s}\hat{t}\hat{u}/(\hat{s}^2+\hat{t}^2+\hat{u}^2)$
as proposed in Ref.~\cite{field}.
However, in our case we found that this choice is equivalent to
$\kappa p_T/z_c$.

At NLO level and no intrinsic transverse momentum Eq.~(\ref{pqcd-def}) is
usually rewritten
with variable change ($x_a$,$x_b$,$z_c$)$\,\to$
($\hat{v}$,$\hat{w}$,$z_c$), where 
$\hat{t}=-(1-\hat{v})\hat{s}$ and $\hat{u}=-\hat{v}\hat{w}\hat{s}$, as
\be
\label{pqcd-vw}
        \frac{\dd\sigma}{\dd y\dd^2p_T} 
        \hspace*{-15mm}&\hspace*{15mm}=\hspace*{-15mm}&\hspace*{15mm} 
        \frac1{\hat{s}}\sum_{abc}
        \int\!\dd\hat{v}\dd\hat{w}
        \frac{\dd z_c}{\pi z_c^2} 
        \ J \\
        & &\dd^2k_{\perp,a}\dd^2k_{\perp,b}\ f_{a/p} f_{b/p} D_{\pi/c}(z_c,Q_f)
        \frac{\dd\sigma^{NLO}}{\dd\hat{v}}  \,\,  ,
        \nonumber 
\ee
with the proper kinematical boundaries, $J$ being the Jacobian of the 
transformation ($1/J=\hat{v}(1-\hat{v})\hat{w}$ for $\k2av$=0), and 
$\dd\sigma^{NLO}/\dd\hat{v}$ is the sum of $2\to2$ and $2\to3$ cross 
sections~\cite{aversa},
\be
 \frac{\dd\sigma^{Born}}{\dd\hat{v}}\delta(1\!-\!\hat{w}) +
 \frac{\alpha_s(Q_r)}{\pi}{\cal K}^{ab\to cd}(\hat{s},\hat{v},\hat{w},Q,Q_r,Q_f)\,\, ,
\ee
with {\em renormalization} scale $Q_r$ (chosen to be equal to the 
factorization scale $Q$). In this paper, however, we use 
Eq.~(\ref{pqcd-def}) directly, since for $\k2av\neq0$ the momentum fraction
$x$ cannot be expressed analytically from $\hat{v}$ and $\hat{w}$, while
the inverse transformation can be done.

Several codes are available for calculating jet cross sections at NLO 
level~\cite{aversa,othernlo}. Here we have chosen to extend the one by 
the Aversa group~\cite{aversa} calculating the partonic cross sections 
at next-to-leading-log level, with the intrinsic transverse momentum 
distribution.
The calculations presented in the following sections were performed
with the MRST-cg PDF~\cite{MRST} and KKP FF~\cite{KKP} parameterization.
 
\section{Results}
\begin{figure}[t]
\centerline{%
\rotatebox{-90}{\includegraphics[height=8.1truecm]%
   {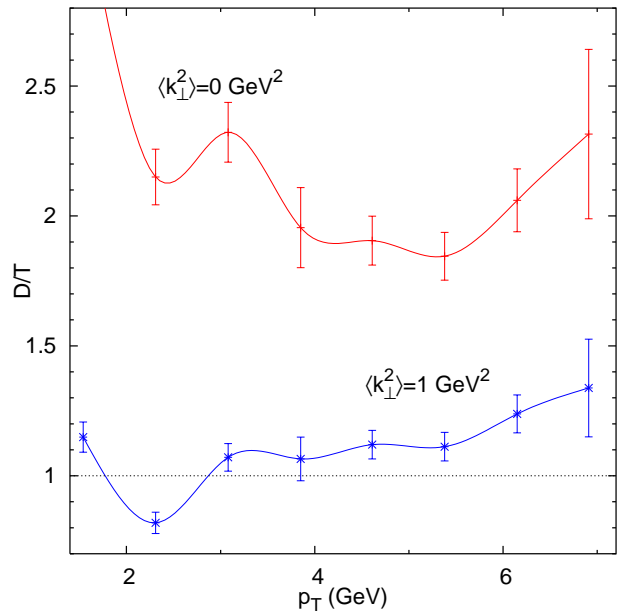}}
}
\caption{Comparison of experimental data~\cite{Antreasyan} to the NLO
pQCD parton model result in $p+p\to\pi^++X$ reaction at $E_{lab}=400$
GeV, with and without intrinsic transverse momentum.}
\end{figure}
\subsection{Comparison to data}
First, we demonstrate the importance of the intrinsic transverse momentum to
reproduce experimental data in the transverse momentum region 
$3$ GeV $< \sqrt{s} < 6$ GeV. In this part, we use the scales proposed in 
Ref.~\cite{aur1} to make a direct comparison to the 400 GeV FNAL
experiment~\cite{Antreasyan}. 
Fig. 1. shows, that at $\sqrt{s}=27.4$ GeV
the NLO calculations of pion production in $pp$ collision
underpredict the experimental data by a factor of
2 using the scale parameters $Q=Q_r=Q_f=p_T/2$ and neglecting
intrinsic transverse momentum (as in~\cite{aur1}).
Decreasing the scales (but still keeping them hard for the applicability
of pQCD) the agreement can be improved, however,
without intrinsic transverse momentum, even if lower scales are
chosen, the data are still underestimated.

The lower line in Fig.~1. presents the NLO calculation {\em with} a partonic
intrinsic transverse momentum distribution of width $\k2av$ =1 GeV$^2$, and
shows a nice agreement with the data.
We note, however, that there is a delicate interplay between the choice of the
scales and the intrinsic transverse momentum $\k2av$, needed to reproduce
the data~\cite{usnew}. 
Typically, increasing the scales increases the value of  $\k2av$.

\subsection{Intrinsic transverse momentum}
\begin{figure}[t]
\centerline{%
{\includegraphics[width=0.49\textwidth,height=8.1truecm]{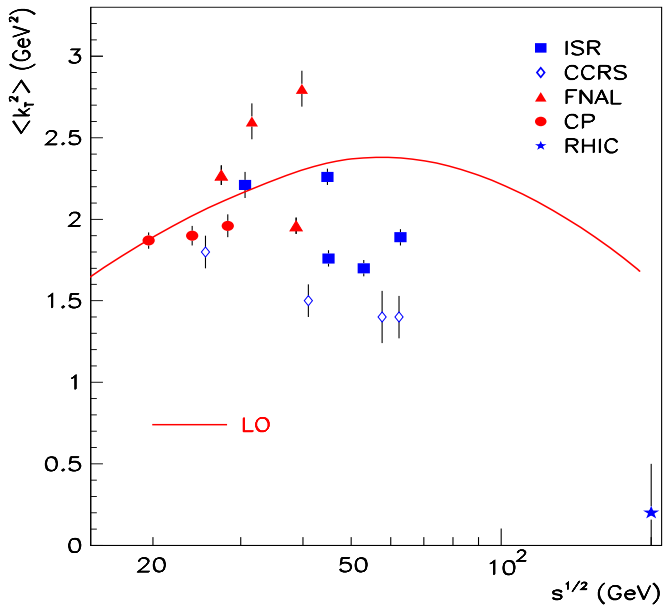}}
}
\caption{Best fits of $\k2av$ to experimental data in $pp\to\pi+X$
reactions from 19.5 to 200 GeV. Different symbols refer to different
experiments, see~\cite{Antreasyan,Exp}. The solid line represents the
average value of a LO calculation~\cite{LOBp}.}
\label{fig-kt2}
\end{figure}

In this section we summarize the results on the intrinsic
transverse momentum width of the partons in the nucleon
fitting the NLO pQCD calculations to the data, generalizing the LO scale
choice of~\cite{BP2002} to $Q=Q_r=\kappa p_T/z_c$, $Q_f=\kappa p_T$,
where $\kappa$ is a ${\cal O}(1)$ number. In this work, we fix
$\kappa=2/3$ at NLO level. Our previous 
study~\cite{BP2002} also showed, that the reproduction of the Cronin peak
in $pA$ collision requires the width of the intrinsic transverse momentum
distribution to be on the order of $\k2av\approx$ 2 GeV$^2$ at energies 
$\sqrt{s}\sim$ 30 GeV, which is achieved by the above choice of scales. A
similar
value of $\k2av$ was extracted from the experimental analysis of jet-angle
distribution~\cite{EXPBROAD}.

Analyzing the $pp\to\pi+X$ experimental data~\cite{Antreasyan,EXPBROAD,Exp},
we deduced the best fit value $\k2av$ for each experiment, similarly to
what is shown in Fig.~1, minimizing the $\chi^2(D/T-1)$ (data over theory) 
ratio 
in the range $3-6$ GeV. The result is presented in Fig.~\ref{fig-kt2},
separately indicating the runs from different experiments and shows a
need for a considerable amount of partonic transverse momentum. We also
checked, that similarly to the LO results~\cite{LOBp}, the extracted width
does not depend on the charge of the pion.


Usually, changing the order of a calculation requires changing of the
underlying scales.
Comparing NLO result to the previous LO ones~\cite{LOBp} one notices
that in order to keep the average transverse momentum width, we had to
increase the scales, dictated by the experimentally obtained Cronin
peak~\cite{BP2002}. Keeping LO scales ($\kappa=1/2$) 
would lead to a substantial
reduction of the width, originating in the mechanism of NLO graphs to
automatically generate transverse momenta. It is remarkable, that
fitting to the nuclear reaction data requires the same width
independent of the order of the calculation used!
\begin{figure}[t]
\centerline{%
\rotatebox{-90}{\includegraphics[height=0.437\textwidth]{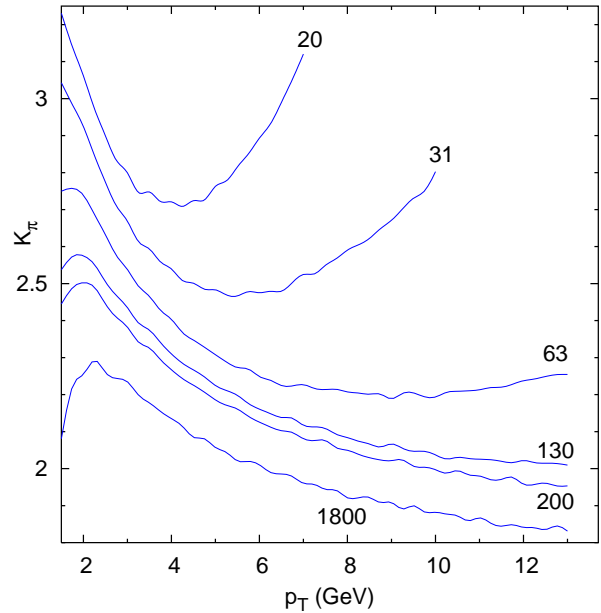}}
}
\caption{$K$ factor at $\k2av=0$ and energies from $\sqrt{s}=20$ GeV 
to 1800 GeV.}
\label{fig-kt0}
\end{figure}

Recent $dAu$ data at RHIC~\cite{dAu} also indicate that at
$\sqrt{s}=200$ GeV more transverse momentum is necessary, than
$\k2av\approx 0-0.5$ GeV$^2$, indicated in Fig.~2. This shows, that the
scale parameter $\kappa$ possibly may also depend on the energy, and a higher,
$\kappa=4/3$ scale with $\k2av=2.5$ GeV$^2$ reproduces well both the
$pp$ and the $dAu$ data without jet quenching effects~\cite{dAuUS}.
The dependence of the scale parameter on the c.m. energy and its
consequences will be studied in a separate paper.
\begin{figure*}
\centerline{%
\rotatebox{-90}{\includegraphics[height=0.9\textwidth]{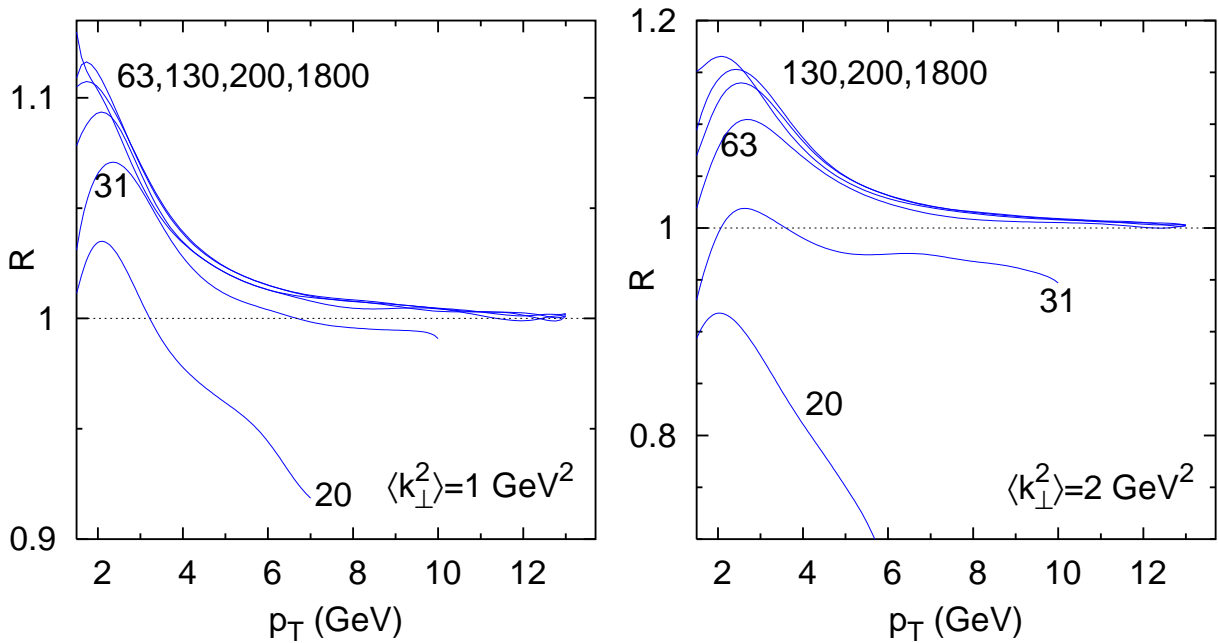}}
}
\caption{Ratio of the pionic $K$ factor at $\k2av=1$ GeV${}^2$ (left) and 
$2$ GeV${}^2$ (right) to $K$ factor at $\k2av=0$ at energies from 
$\sqrt{s}=20$ GeV to 1800 GeV.}
\label{fig-kt}
\end{figure*}

\subsection{$K$ factor}

Since most calculations (especially, for nucleus-nucleus collisions)
are still based on LO, it is useful to provide a well-founded $K$ factor
for these faster calculations. Fig.~\ref{fig-kt0} shows the ratio of the
full NLO calculation to the Born term with no intrinsic transverse momentum.
Indeed, at high energy and transverse momenta $K_\pi$ approaches the
well known value of 2, however, in the low transverse momentum region
it has a strong $p_T$ dependence. As a first approximation, for
$\sqrt{s}\gtrsim$ 60 GeV, the pionic $K$ factor can be taken energy
independent~\cite{JPG}.

Since the intrinsic transverse momentum may have a different effect on the
Born term than on the higher order (HO) processes, the latter contributing the
dominant part, it is worth to study the dependence of the $K$ factor on the
width $\k2av$. We present this behavior in Fig.~\ref{fig-kt}, indicating
the ratio $R=K(\k2av)/K(0)$ at $\k2av=$ 1 and 2 GeV$^2$ (with 
$\kappa=2/3$ scale parameter) and at different energies
from $\sqrt{s}=$ 20 to 1800 GeV. While at low energies the ratio $R$ shows
a large decrease with increasing $\k2av$ (mainly due to the efficiency of the
intrinsic transverse momentum at low energies in enhancing the Born term faster
than the HO corrections), from $\sqrt{s}\gtrsim$ 60 GeV its value is
always above 1, indicating, that HO corrections are gaining more from
transverse momentum, than the Born term. The maximal enhancement is
peaked at $p_\perp\approx 2-3$ GeV, with 10\% correction at
$\k2av=1$ GeV$^2$ and 20\% correction at $\k2av=2$ GeV$^2$, while
for $p_T\gtrsim$ 4 GeV, the transverse momentum driven correction vanishes,
validating LO calculations
in this range with $\k2av$ independent $K$ factors~\cite{Wong,BP2002}.

\section{Conclusions}
In this paper we introduced initial transverse momentum distributions
into the parton-based description of hadron production in NLO level
pQCD calculations and demonstrated, that such an extension is necessary
even in NLO to reproduce the $pp$ experimental data.
From the analysis of most experiments in the 3 GeV $\lesssim p_T \lesssim$
6 window a width of intrinsic transverse momentum distribution on the
order of  $\k2av\approx\ 2$ GeV$^2$ was fitted.
The obtained  precision of the description of pion production in $pp$
collisions is high enough to find possible collective effects in nuclear
collisions.

We presented the pionic $K$-factor (full NLO cross section to the Born
term) for several energies and transverse momentum values, and found a
pronounced dependence on $p_T$ in the 3 GeV $\lesssim p_T\lesssim$ 6
window, where nuclear effects show up most prominently.
Furthermore, we investigated in detail the modifications of the Born and
higher order terms due to the presence of an intrinsic transverse momentum
and concluded, that above certain energy in the above mentioned $p_T$
window the higher order contribution raises faster, than the
Born term and hence the $K$-factor increases compared to its value without
intrinsic transverse momentum. For $p_T\gtrsim$ 4 GeV this enhancement may
be neglected, justifying $\k2av$ independent $K$-factor
calculations, however, at smaller transverse momenta
the correction goes up to 20\%, just in the order of the
measured nuclear enhancement. The detailed
study of the nuclear enhancement, and its role in fixing the scales of an
NLO calculation will be carried out separately.

\section*{Acknowledgements}

We thank G. David for many useful discussions.
This work was supported in part by  U.S. DOE grant DE-FG02-86ER40251, NSF grant
INT-0000211, and Hungarian grants  FKFP220/2000, OTKA-T034842 and OTKA-T43455.
The support of the Bergen Computational Physics Laboratory in the
framework of the European Community - Access to Research Infrastructure 
action of the Improving Human Potential Programme is gratefully acknowledged.

\end{document}